\documentclass[10pt,twocolumn]{revtex4-2} 
\usepackage[dvips]{color}
\usepackage{epsfig}
\usepackage{amssymb}
\usepackage{latexsym}
\usepackage{graphicx}
\begin{document}

\title{Monte Carlo simulations of phonon propagation in the Fermi-sea of Weyl spinors\\ and detection of hysteresis effects using Groupoids}
\author
{ Sadataka Furui}\affiliation{(Formerly) Graduate School of Science and Engineering, Teikyo University, Utsunomiya, Japan }
 \email{furui@umb.teikyo-u.ac.jp}
\author
{ Serge Dos Santos}\affiliation{INSA Centre\, Val de Loire, UMR 1253 Imaging and Brain : iBrain, Universit\'e de Tours, Blois Cedex, France }
\email{serge.dossantos@insa-cvl.fr} 
\date{\today }

\begin{abstract}
We considered symplectic quaternions sitting on a $(2+1)D$ lattice in momentum space which interact with nearest neighbor interactions.  The action is expressed by fixed point actions given by the superposition of blocks whose length of the contour is 4, 6 or 8 lattice spacings.
 
The spinors are on a finite $2D$ plane expressed as $u_1 a e_1+ u_2 a e_2$ (A type) and/or on two $2D$ planes separated by $\pm a e_1\wedge e_2$ (B type). 
In the A type, link actions on counterclockwise rotating loop and on clockwise rotating loop cancel, however in the B type, the direction of links   
between the lower $2D$ plane and the upper $2D$ plane induces differences in the action.  The time-reversal conserving and spin rotational symmetry breaking actions can be measured. Result of Monte-Carlo simulations of $64\times 64$,
$128\times 128$ and $256\times 256$ spatial lattices are presented.
\end{abstract}
\maketitle
\section{ Introduction}
In order to improve the resolution of the image produced in non-destructive testings(NDT), it is inportant to take into account hysteresis effects of materials that allow propagation of solitonic phonons.

In the NDT, transducers and receivers are placed on a $2D$ plane and the convolution of a phonon and/or a time reversed phonon scattered by a singularity in the material is measured\cite{LSDS17}.  The $2D$ plane is expanded by unit vectors ${\bf e}_1$ and ${\bf e}_2$, and ithe Clifford algebra ${\mathcal C}\ell_2$, an arbitrary element is written as $u=u_0+u_1{\bf e}_1+u_2{\bf e}_2+u_{12}{\bf e}_{12}$ where ${\bf e}_{12}$ is a bivector\cite{Garling11,Lounesto01,Porteous95}.

For ${\bf a}=a_1{\bf e}_1+a_2{\bf e}_2$ and ${\bf b}=b_1{\bf e}_1+b_2{\bf e}_2$,
\begin{eqnarray}
{\bf a}\cdot{\bf b}&=&a_1 b_1+a_2 b_2,\nonumber\\
{\bf a}\wedge {\bf b}&=&(a_1 b_2-a_2 b_1){\bf e}_{12}.\nonumber 
\end{eqnarray}

 We considered the fixed point Wilson actions in one loop adopted by DeGrand et al.\cite{DGHHN95}. They considered in the $(3+1)D$ lattice, 28 Fixed point (FP) actions of length less than or equal to 8 lattice lengths. We classify the FP actions to  A type which consist of loops on one $2D$  plane expanded by $e_1$ and $e_2$, and B type which consist of loops on two parallel $2D$ planes connected by two links in the direction $e_1\wedge e_2$ and in the direction $e_2\wedge e_1$.

In the $(2+1)D$ we consider the following 20 loops
\begin{itemize}
\item A: Loops $L1,$ $L2,$ $L5,$ $L6,$ $L11,$ $L12,$ $L18$.
\item B: Loops $L3,$ $L4,$ $L7,$ $L8,$ $L9,$ $L10,$ $L13,$ $L14,$ $L15,$ $L16,$ $L17,$ $L26,$ $L27$.
\end{itemize}

The $L19,\cdots ,L25$ and $L28$ used in \cite{DGHHN95} are irrelevant in our $(2+1)D$ model.

Presence of Fermi sea in the propagation media makes the treatment of the Schwinger function which depends on the Fermi momentum $p_F$ and effective mass $m$.

 The structure of this presentation is as follows. In sect. II, we review the Monte Carlo simulation of fixed point actions.
 In Sec. III we present results of Monte-Carlo search of minimal Wilson action of a model of Weyl spinors sitting on $(2+1)D$ lattices.

 In Sec. IV, a perspective of detecting hysteresis effects using groupoids is presented and discuss about timereversal symmetry preserved and spin rotation symmetry broken phase.
 
 Summary and Conclusion are given in Sec.V.
 
 \section{Monte Carlo simulation of fixed point actions}
  In this section, we review the strategies for getting the minimal action as a linear combination of plaquette actions for A-type loops and plaquette actions and an average of link action of counterclockwise rotating and that of clockwise rotating loop.
 
In the case of A-type loops, we first produce 7 random numbers $randa[i], i=0,\cdots 6$ and make 12 ordered sets $randb[i]$ as shown in \cite{FDS21}. We store eigen values of the loops 
$L1,$ $L2,$ $L5,$ $L6,$ $L11,$ $L12,$ $L18$, which are denoted as $\lambda_n$ for the loop $L_n$ $(n=1,2,\cdots,18)$ at the lattice $(u_1,u_2)$.  In this work we take $u_i=0, \frac{1}{N_{max}}\Delta u,\cdots, \Delta u$ $(i=1,2) $ and $N_{max}=4,8$ and 16.  On the torus $T^2, 16\times N_{max}$ spatial lattice points are considered and  actions represented by quaternions are calculated by using Mathematica\cite{Wolfram20}.

The total plaquette action is in the form
\begin{eqnarray}
S_{PA}[i,j]&=&c_1 [\lambda_1(i,j)-lim_1]+c_2 [\lambda_2(i,j)-lim_2]\nonumber\\
&&+c_5 [\lambda_5(i,j)-lim_5]+c_6[\lambda_6(i,j)-lim_6]\nonumber\\
&&+c_{11} [\lambda_{11}(i,j)-lim_{11}]+c_{12}[\lambda_{12}(i,j)-lim_{12}]\nonumber\\&&+c_{18} [\lambda_{18}(i,j)-lim_{18}],\nonumber
\end{eqnarray}
where  in the case of $\Delta u=\frac{1}{4}$
\begin{eqnarray}
&&lim_1=2.71,\quad lim_2=2.83,\quad lim_5=10.97,\nonumber\\
&&lim_6=5.54,\quad lim_{11}=11.0,\quad lim_{12}=5.56,\nonumber\\
&&lim_{18}=4.26,\nonumber
\end{eqnarray}
are evaluated graphically.

On the computer SQUID at Osaka University, we calculate 
\begin{eqnarray}
 sp0_n[j]& =&\sum_i (\lambda_n[0][j] - lim\lambda_n) * randa[i];\nonumber\\
   sp1_n[j]&=&\sum_i(\lambda_n[1][j] - lim\lambda_n) * randa[i];\nonumber\\
   sp4_n[j]& =&\sum_i (\lambda_n[4][j] - lim\lambda_n) * randa[i];\nonumber
\end{eqnarray}
and plot
\[
\sum_{n=1,2,\cdots,18} sp0_n[j], 
\sum_{n=1,2,\cdots,18} sp1_n[j],
\sum_{n=1,2,\cdots, 18} sp4_n[j]
\]
as a function of  $u_j=\frac{j}{4}$, ($j=0,1,\cdots,64$)..

The plaquette actions of A type loops are
\begin{figure}[htb]
\begin{center}
\includegraphics[width=5cm,angle=0,clip]{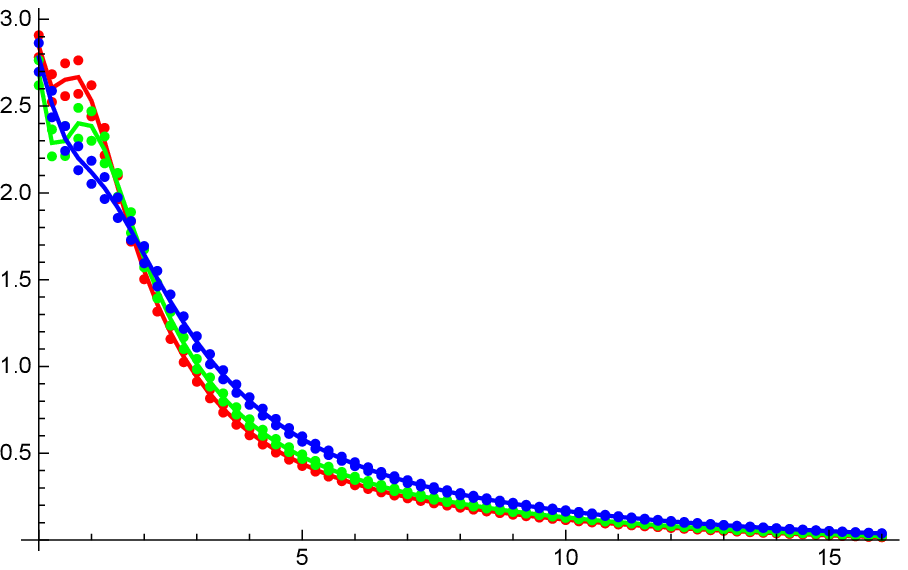}
\end{center}
\end{figure}
\begin{figure}[htb]
\begin{center}
\includegraphics[width=5cm,angle=0,clip]{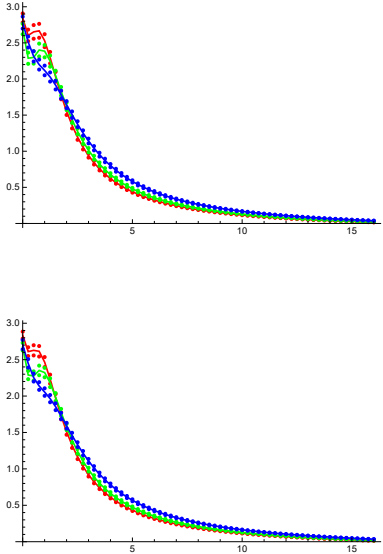}
\end{center}
\caption{ Plaquette actions of A type loops (Two sets of 7 random numbers with 12 different orderings.) The ordinate is $u_1=\frac{j}{4}\Delta u$ ($u_2=0$(red), $u_2=\frac{1}{4}\Delta u$(green), $u_1=\Delta u$(blue). }
\label{splaq}
\end{figure}

 The link actions of A-type loops are in the form
 \begin{eqnarray}
S^e_{LA}[i,j]&=&c_1 \ell_1(i,j)+c_2 \ell_2(i,j)\nonumber\\
&&+c_5\ell_5(i,j)+c_6\ell_6(i,j)\nonumber\\
&&+c_{11}\ell_{11}(i,j)+c_{12}\ell_{12}(i,j)+c_{18}\ell_{18}(i,j),\nonumber
\end{eqnarray}
where 
\begin{eqnarray}
&&\ell_1(i,j)=dss1[i/2,j/2]+dss2[1/8+i/2,j/2]\nonumber\\
&&-dss2[1/8+i/2,1/8+j/2]-dss1[1/8+i/2,j/2].\nonumber
\end{eqnarray}
We remark the upper right corner $2\times 2$ matrices of $dss1[u_1,u_2]$ and $dss2[u_1,u_2]$ are \cite{SF21a,SF21c}
\begin{eqnarray}
&&dss1[u_1,u_2]_{13}=\frac{2u_1}{1+|u|^2}+\sqrt{-1}[-\frac{2u_2}{1+|u|^2}+\sqrt{-1}\frac{1-|u|^2}{1+|u|^2}]\nonumber\\
&&dss1[u_1,u_2]_{14}=\frac{2u_1}{1+|u|^2}+\sqrt{-1}[\frac{2u_2}{1+|u|^2}+\sqrt{-1}\frac{1-|u|^2}{1+|u|^2}]\nonumber\\
&&dss1[u_1,u_2]_{23}=-\frac{2u_1}{1+|u|^2}-\sqrt{-1}[-\frac{2u_2}{1+|u|^2}+\sqrt{-1}\frac{1-|u|^2}{1+|u|^2}]\nonumber\\
&&dss1[u_1,u_2]_{24}=\frac{2u_1}{1+|u|^2}-\sqrt{-1}[-\frac{2u_2}{1+|u|^2}-\sqrt{-1}\frac{1-|u|^2}{1+|u|^2}]\nonumber\\
&&dss2[u_1,u_2]_{13}=-\frac{2\sqrt{-1} u_1}{1+|u|^2}+\frac{2u_2}{1+|u|^2}+\sqrt{-1}\frac{1-|u|^2}{1+|u|^2}\nonumber\\
&&dss2[u_1,u_2]_{14}=-\frac{2\sqrt{-1} u_1}{1+|u|^2}-\frac{2u_2}{1+|u|^2}-\sqrt{-1}\frac{1-|u|^2}{1+|u|^2}\nonumber\\
&&dss2[u_1,u_2]_{23}=-\frac{2\sqrt{-1} u_1}{1+|u|^2}+\frac{2u_2}{1+|u|^2}-\sqrt{-1}\frac{1-|u|^2}{1+|u|^2}\nonumber\\
&&dss2[u_1,u_2]_{24}=\frac{2\sqrt{-1} u_1}{1+|u|^2}+\frac{2u_2}{1+|u|^2}-\sqrt{-1}\frac{1-|u|^2}{1+|u|^2},\nonumber
\end{eqnarray}
where $|u|^2=u_1^2+u_2^2$.

The counterclockwise rotating and clockwise rotating loop contributions cancel with each other.

In the case of B-type loops, we first produce 13 random numbers $randb[i], i=0,\cdots 13$ and make 50 ordered sets as shown in \cite{FDS21}. We store eigen values of the loops $L3,$ $L4,$ $L7,$ $L8,$ $L9,$ $L10,$ $L13,$ $L14,$ $L15,$ $L16,$ $L17,$ $L26,$ $L27$, which are denoted as $\lambda_n$ for the loop $L_n$ $(n=3,4,\cdots,27)$. 

\begin{eqnarray}
S_{PB}[i,j]&=&c_3 [\lambda_3(i,j)-lim_3]+c_4 [\lambda_4(i,j)-lim_4]\nonumber\\
&&+c_7 [\lambda_7(i,j)-lim_7]+c_8[\lambda_8(i,j)-lim_8]\nonumber\\
&&+c_9 [\lambda_9(i,j)-lim_9]+c_{10}[\lambda_{10}(i,j)-lim_{10}]\nonumber\\
&&+c_{13} [\lambda_{13}(i,j)-lim_{13}]+c_{14}[\lambda_{14}(i,j)-lim_{14}]\nonumber\\
&&+c_{15} [\lambda_{15}(i,j)-lim_{15}+c_{16} [\lambda_{16}(i,j)-lim_{16},\nonumber\\
&&+c_{17} [\lambda_{17}(i,j)-lim_{17}+c_{26} [\lambda_{26}(i,j)-lim_{26},\nonumber\\
&&+c_{27} [\lambda_{27}(i,j)-lim_{27}\nonumber
\end{eqnarray}
 where in the case of $\Delta u=\frac{1}{4}$,
\begin{eqnarray}
&&lim_3=5.41,\quad lim_4=5.41,\quad lim_7=7.64,\nonumber\\
&&lim_8=10.81,\quad lim_9=10.79,\quad lim_{10}=7.66,\nonumber\\
&&lim_{13}=7.90,\quad lim_{14}=5.67, \quad lim_{15}=7.91,\nonumber\\
&&lim_{16}=7.63,\quad lim_{17}=5.67, \quad lim_{26}=11.01\nonumber\\
&&lim_{27}=11.02.\nonumber
\end{eqnarray}

Numerically we calculate
\begin{eqnarray}
 &&sp0_n[j] =\sum_i (\lambda_n[0][j] - lim\lambda_n) * randb[i];\nonumber\\
 &&sp1_n[j] =\sum_i (\lambda_n[1][j] - lim\lambda_n) * randb[i];\nonumber\\
 &&sp4_n[j] =\sum_i (\lambda_n[4][j] - lim\lambda_n) * randb[i];\nonumber
\end{eqnarray}
and make summations
\[
\sum_{n=3,4,\cdots,27}sp0_n[j], \sum_{n=3,4,\cdots,27}sp1_n[j], \sum_{n=3,4,\cdots,27}sp4_n[j]
\]
Plaquette actions of two samples are shown in Fig.\ref{splaq}.

We compare for example the path of $L3$: 
\begin{eqnarray}
&&{\bf u}\to {\bf u}+\frac{1}{8}e_1\to {\bf u}+\frac{1}{8}e_1+\frac{1}{8}e_2\to\nonumber\\
&& {\bf u}+\frac{1}{8}e_1+\frac{1}{8}e_2+\frac{1}{8}e_1\wedge e_2\to {\bf u}+\frac{1}{8}e_1+\frac{1}{8}e_1\wedge e_2\to\nonumber\\
&& {\bf u}+\frac{1}{8}e_1\wedge e_2\to {\bf u},\nonumber
\end{eqnarray}
and the path of
 \begin{eqnarray}
&&{\bf u}\to {\bf u}+\frac{1}{8}e_1\to {\bf u}+\frac{1}{8}e_1+\frac{1}{8}e_2\to\nonumber\\
&& {\bf u}+\frac{1}{8}e_1+\frac{1}{8}e_2-\frac{1}{8}e_1\wedge e_2\to {\bf u}+\frac{1}{8}e_1-\frac{1}{8}e_1\wedge e_2\to\nonumber\\
&& {\bf u}+\frac{1}{8}e_1\wedge e_2\to {\bf u},\nonumber
 \end{eqnarray}
 which we call $e$ link and $f$ link, respectively.

The counterclockwise rotating loop's link actions are
\begin{eqnarray}
S_{LB}[i,j]&=&c_3\ell_3(i,j)+c_4\ell_4(i,j)+c_7\ell_7(i,j)\nonumber\\
&&+c_8\ell_8(i,j)+c_9\ell_9(i,j)+c_{10}\ell_{10}(i,j)\nonumber\\
&&+c_{13}\ell_{13}(i,j)+c_{14}\ell_{14}(i,j)\nonumber\\
&&+c_{15}\ell_{15}(i,j)+c_{16}\ell_{16}(i,j)+c_{17}\ell_{17}(i,j)\nonumber\\
&&+c_{26}\ell_{26}(i,j)+c_{27}\ell_{27}(i,j).\nonumber
\end{eqnarray}

\begin{figure}[htb]
\begin{center}
\includegraphics[width=4cm,angle=0,clip]{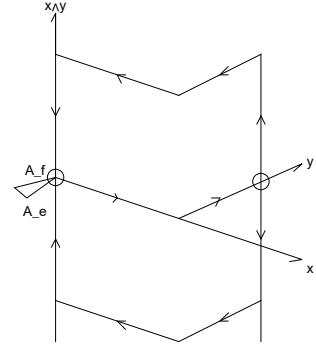}
\end{center}
\caption{ The  link actions of counterclockwise rotating and clockwise rotating $L3$ loop.  }
\label{deltaA1}
\end{figure}

Here the counterclockwise rotating link action of $L3$ is
\begin{eqnarray}
&&\ell_3^f(i,j) =dss1[i,j]+dss2[1/8+i,j]+dss3[1/8+i,1/8+j]\nonumber\\
&&-dss2[1/8+i,1/8+j]-dss1[1/8+i,j]-dss3[1/8+i, j],\nonumber
\end{eqnarray}
and clockwise rotating link action of $L3$ is
\begin{eqnarray}
&&\ell_3^e(i,j) =dss1[i,j]+dss2[1/8+i,j]-dss3[1/8+i,1/8+j]\nonumber\\
&&-dss2[1/8+i,1/8+j]-dss1[1/8+i,j]+dss3[1/8+i, j].\nonumber
\end{eqnarray}
The $L3$ link of counterclockwise rotating and clockwise rotating are shown in Fig.\ref{deltaA1}.

Expressions for other loops are given in \cite{SF21a,SF21c,SF21b}.

 The upper right corner of the $2\times 2$ matrix of $dss3[i,j]$ is
\begin{eqnarray}
&&dss3[u_1,u_2]_{13}=\frac{(2-2\sqrt{-1})u_1}{1+|u|^2}+\frac{2u_2}{1+|u|^2}+\sqrt{-1}\frac{1-|u|^2}{1+|u|^2}\nonumber\\
&&+\sqrt{-1}[-\frac{2u_2}{1+|u|^2}+\sqrt{-1}\frac{1-|u|^2}{1+|u|^2}]\nonumber\\
&&dss3[u_1,u_2]_{14}=\frac{(2-2\sqrt{-1})u_1}{1+|u|^2}-\frac{2u_2}{1+|u|^2}-\sqrt{-1}\frac{1-|u|^2}{1+|u|^2}\nonumber\\
&&+\sqrt{-1}[-\frac{2u_2}{1+|u|^2}+\sqrt{-1}\frac{1-|u|^2}{1+|u|^2}]\nonumber
\end{eqnarray}
\begin{eqnarray}
&&dss3[u_1,u_2]_{23}=-\frac{(2+2\sqrt{-1})u_1}{1+|u|^2}+\frac{2u_2}{1+|u|^2}-\sqrt{-1}\frac{1-|u|^2}{1+|u|^2}\nonumber\\
&&-\sqrt{-1}[-\frac{2u_2}{1+|u|^2}+\sqrt{-1}\frac{1-|u|^2}{1+|u|^2}]\nonumber\\
&&dss3[u_1,u_2]_{24}=\frac{(2+2\sqrt{-1})u_1}{1+|u|^2}+\frac{2u_2}{1+|u|^2}-\sqrt{-1}\frac{1-|u|^2}{1+|u|^2}\nonumber\\
&&-\sqrt{-1}[-\frac{2u_2}{1+|u|^2}-\sqrt{-1}\frac{1-|u|^2}{1+|u|^2}]\nonumber
\end{eqnarray}
The trace $dss3[u_1,u_2]_{13}+dss3[u_1,u_2]_{24}$ becomes
\[
\frac{4u_1+4u_2-2(1-|u|^2)}{1+|u|^2}
\] 
depends on the position of the two links that connect between the two projected $2D$ complex planes defined by $(u_1,u_2)$.

The lattice spacing of $u_1$ and $u_2$ are defined to be the same $\Delta u$, and we consider cases of $\frac{1}{4}, \frac{1}{8}$ and $\frac{1}{16}$.

\begin{figure*}[htb]
\begin{minipage}[b]{0.47\linewidth}
\begin{center}
\includegraphics[width=5cm,angle=0,clip]{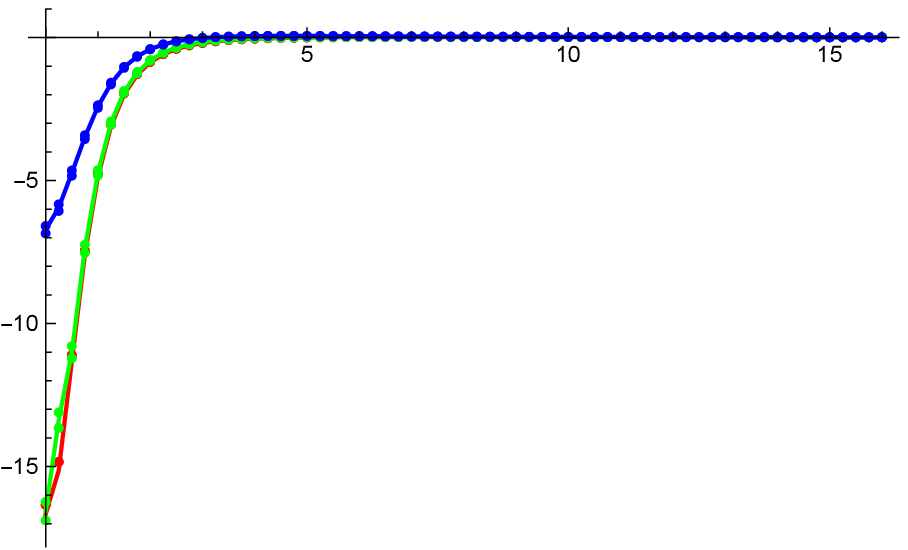}
\end{center}
\end{minipage}
\hfill
\begin{minipage}[b]{0.47\linewidth}
\begin{center}
\includegraphics[width=5cm,angle=0,clip]{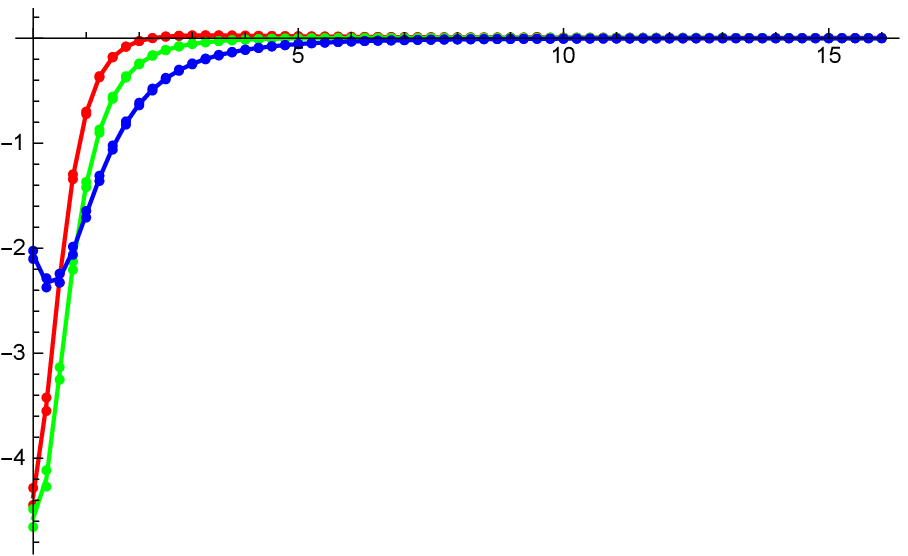}
\end{center}
\end{minipage}
\caption{ Link actions of B type loops (A set using 13 random number parameters  with 50 different orderings). $e-\alpha$ (left) and $f-\alpha$ (right). The ordinate is $u_2=\frac{j}{4}\Delta u$ ($j=0,1,\cdots,64$). $u_2=0$(red), $u_1=\frac{1}{4}\Delta u$(green), $u_1=\Delta u$(blue).  }
\label{link}
\end{figure*}

 The Fig.\ref{link} show the Monte-Carlo expectation value of $S_{LB}^e[i,j]$ and $S_{LB}^f[i,j]$ for fixed $u_1=0, \frac{1}{4},1$ and $u_2=\frac{j}{4}$ ($j=0,\cdots,64$) 

\begin{figure*}
\begin{minipage}[b]{0.47\linewidth}
\begin{center}
\includegraphics[width=5cm,angle=0,clip]{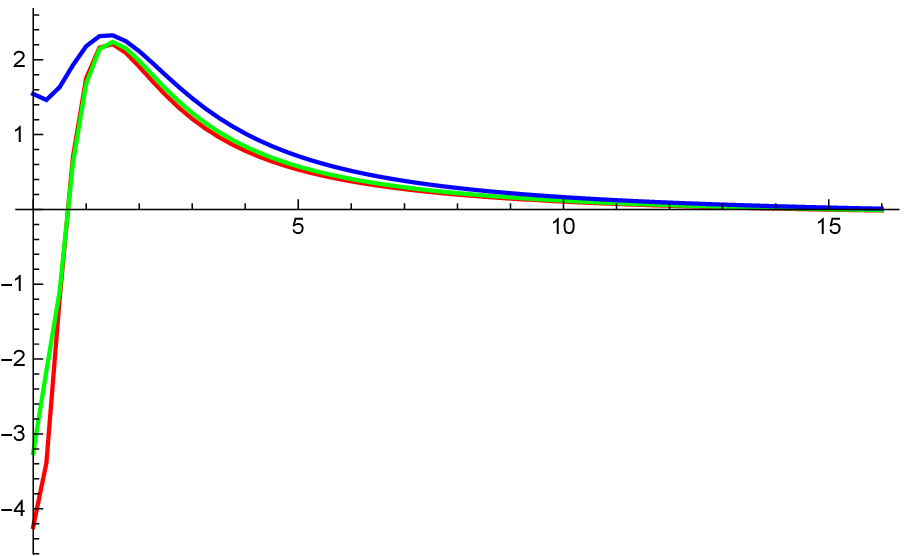}
\end{center}
\end{minipage}
\hfill
\begin{minipage}[b]{0.47\linewidth}
\begin{center}
\includegraphics[width=5cm,angle=0,clip]{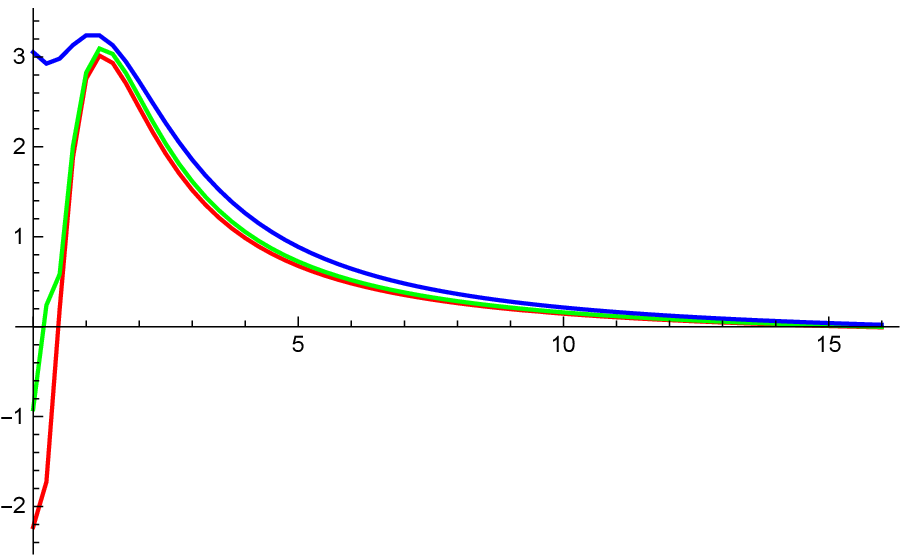}
\end{center}
\end{minipage}
\caption{ The sum of plaquette actions and the average of link actions of counterclockwise rotating and clockwise rotating loops of B type. The sample $\alpha$ (left)  and the sample $\beta$ (right). }
\label{splaq}
\end{figure*}

\begin{figure*}[htb]
\begin{minipage}[b]{0.47\linewidth}
\begin{center}
\includegraphics[width=7cm,angle=0,clip]{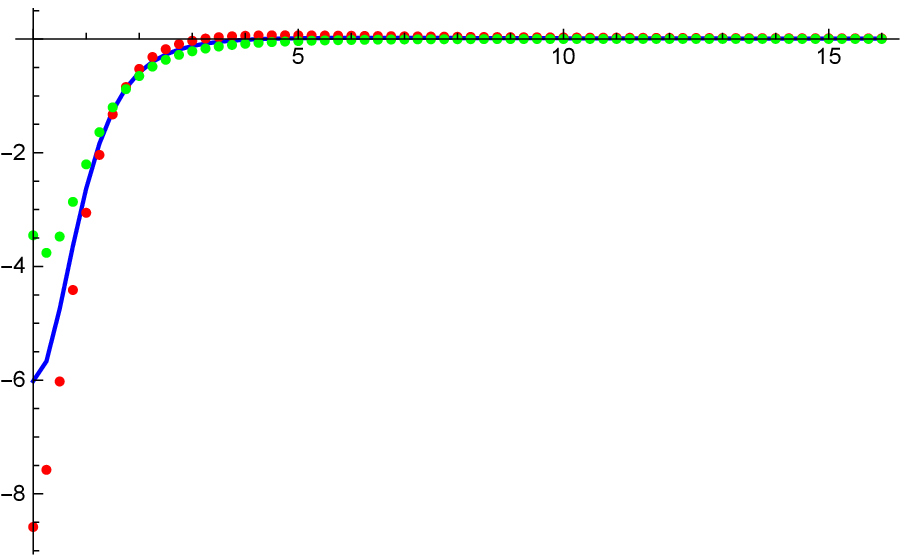} 
\end{center}
\end{minipage}
\hfill
\begin{minipage}[b]{0.47\linewidth}
\begin{center}
\includegraphics[width=7cm,angle=0,clip]{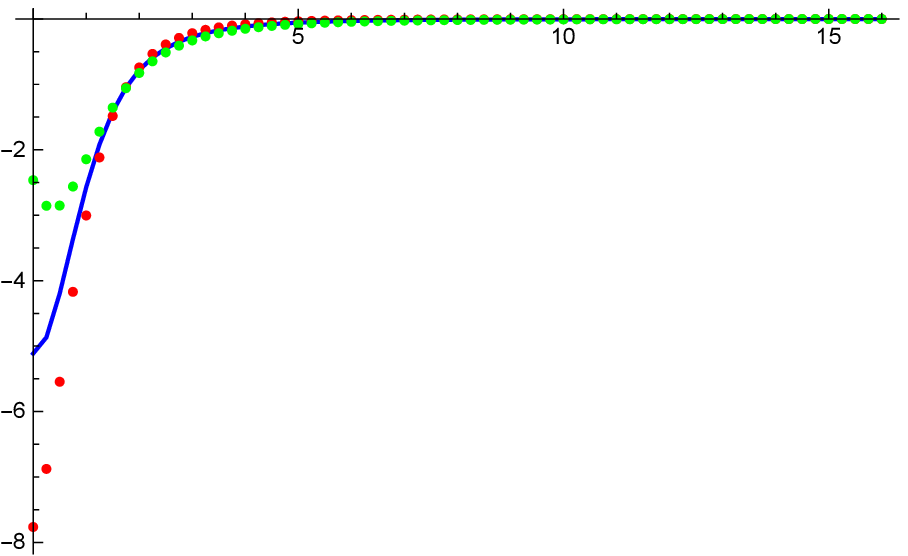} 
\end{center}
\end{minipage}
\caption{  $B$ type link actions at $u_2=\Delta u$ as a function of $u_1$ of the sample $\gamma$(left) and the sample $\beta$(right). Average of $f$link and $e$link is solid blue, $f$link is dotted red,
$e$link is dotted green. Difference of $e$ and $f$ is small in sample $\gamma$. }
\end{figure*}

The study is extended to
 $u_j=\frac{j}{8}\Delta u, (j=0,1,\cdots,128)$. and $u_j=\frac{j}{16}\Delta u, (j=0,1,\cdots,256)$.

\section{Results of Monte-Carlo simulations}
We performed Monte Carlo(MC) simulations of the weight parameters $c_1,c_2,c_5,c_6,c_{11},c_{12},c_{18}$  of the A-type fixed point actions and $c_3,c_4,c_7,c_8,c_9,c_{10},c_{13}, c_{14},c_{15},c_{16},c_{17}, c_{26}, c_{27}$ of the B-type fixed point action, using the pseudo random number generators of Mathematica\cite{Wolfram20} and that of SQUID supercomputer at the Osaka University.

In the $2D$ momentum space, the Wilson link action of Weyl spinors at high $u$ tends to zero and are be asymptotically free, but the plaquette actions tends to non-zero constant. 
 
The A-type link action of counterclockwise rotating loops and clockwise rotating loops in our model cancel, and only the plaquette actions need to be considered to fix the 7 weight parameters. By the choice of fixed point actions, statistical variances are not so large except the infrared region, where both $u_1$ and $u_2$ are small. The plaquette action $S_{PA}$ have similar dependence on $u_2$,

The B-type actions depend on the direction of bivectors ${\bf e}_1\wedge {\bf e}_2$ that connect the two $2D$ planes on which the wave packet of a phonon propagate.

As pointed out in \cite{BG95}, sound waves in fermionic media have effective mass and via careful analysis of the convolution of the ordinary sound and time reversed sound wave, there exists possibility of detecting the gravitational effects. 

We observed large random number dependence in the actions $S^e_{LB}$ and $S^f_{LB}$, but the dependence in the average of $S^e_{LB}$ and $S^f_{LB}$ is not large.

Although there are uncertainties in the relative weight of $S_{LB}$ and $S_{PB}$, the sum of $S_{LB}$ and $S_{PB}$ have similar dependence on $u_2$, and proper weight parameters may be obtained by accumulating large number of random parameter samples.

 In the case of $\Delta u=\frac{1}{4}$, the total action of the sample $\beta$, showed  a zigza behaviors at around $u_2=\frac{1}{4}\Delta u$, and in the case of $\Delta u=\frac{1}{8}$, singular behaviors for certain $u_2$ were observed in the average link action as a functiion of $u_1$.   

The absolute value of the Fourier transform of A type plaquette action of $ u_2=\frac{7}{8}\Delta u$ and $u_2=\frac{8}{8}\Delta u$ of the sample $\alpha$ and the sample $\beta$have similar structure, but the role of $u_2=\frac{7}{8}\Delta u$ and that of $u_2=\frac{8}{8}\Delta u$ are interchanged. An oscillating spectrum becomes tangential to the other smooth spectrum.  

 We compared the spectrum of sample $\gamma$, $u_2=\frac{15}{16}\Delta u$ and $u_2=\frac{16}{16}\Delta u_2$. 

\begin{figure*}[htb]
\begin{minipage}[b]{0.47\linewidth}
\begin{center}
\includegraphics[width=7cm,angle=0,clip]{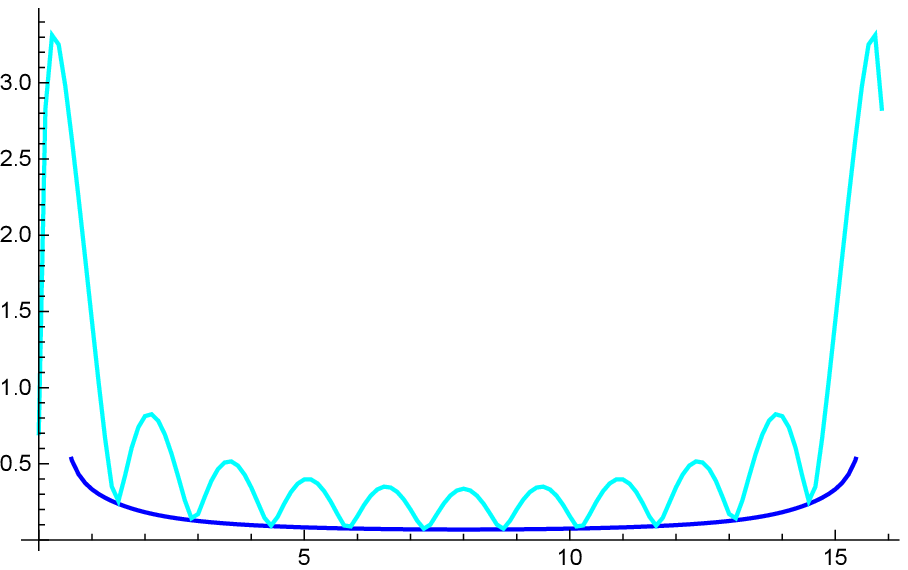} 
\end{center}
\end{minipage}
\hfill
\begin{minipage}[b]{0.47\linewidth}
\begin{center}
\includegraphics[width=7cm,angle=0,clip]{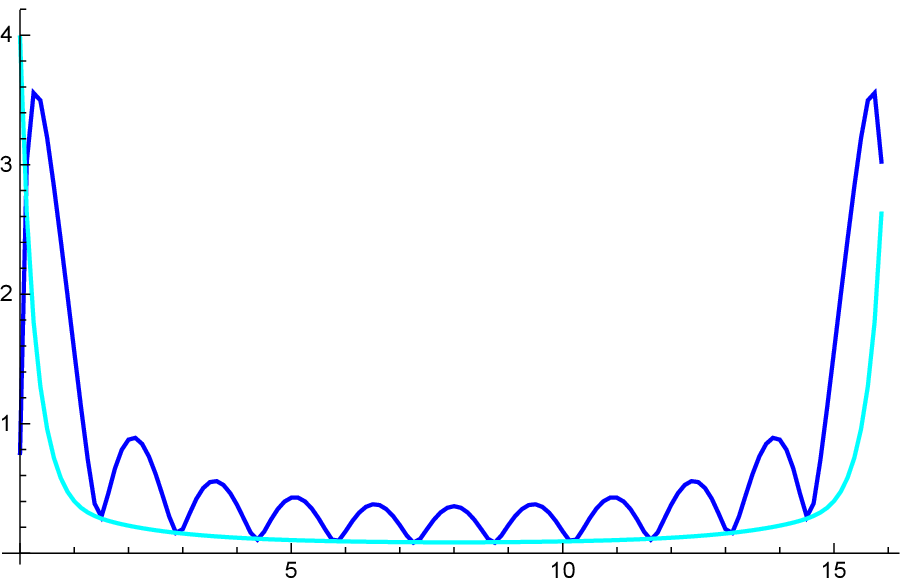} 
\end{center}
\end{minipage}
\caption{Absolute value of the Fourier transform of A type plaquette actions as a function of $u_1$. Pale blue line corresponds to $u_2=\frac{7}{8}\Delta u$, Blue line corresponds to $u_2=\Delta u$.  The sample $\alpha$(left) and the sample $\beta$(right) are diferrent in their 7 random numbers.}
\end{figure*}
The actions of $e$ link and $f$ link have large difference in samples $\alpha,\beta$ and $\delta$.
 When $u_2=\frac{7}{8}\Delta u$, the sample $\gamma$ whose difference of $e$ link and $f$ link is small, show exceptionally one peak, but the other samples show two peakes. In NDT, signals like that of $\gamma$ sample should be used for fitting the weight of Wilson loops.

\begin{figure}[htb]
\begin{center}
\includegraphics[width=4cm,angle=0,clip]{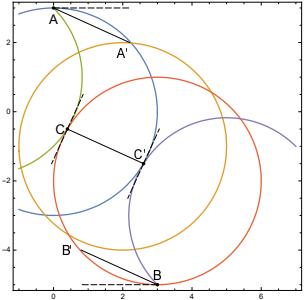}
\end{center}
\caption{ Groupoid multiplications $h\sim AC, g\sim CB, gh\sim AB$ and $h'\sim A'C', g'\sim C'B, g'h'\sim A'B\sim A'B$ . }
\label{holonomy}
\end{figure}

 In the case of A type plaquette actions, the spectrum of $u_1$ as a function of $u_2=\frac{N}{16}\Delta u$ do not depend on whether $N$ is even or odd except large $u_2$ region shown by blue and pale blue curves.

\begin{figure*}[htb]
\begin{minipage}[b]{0.47\linewidth}
\begin{center}
\includegraphics[width=7cm,angle=0,clip]{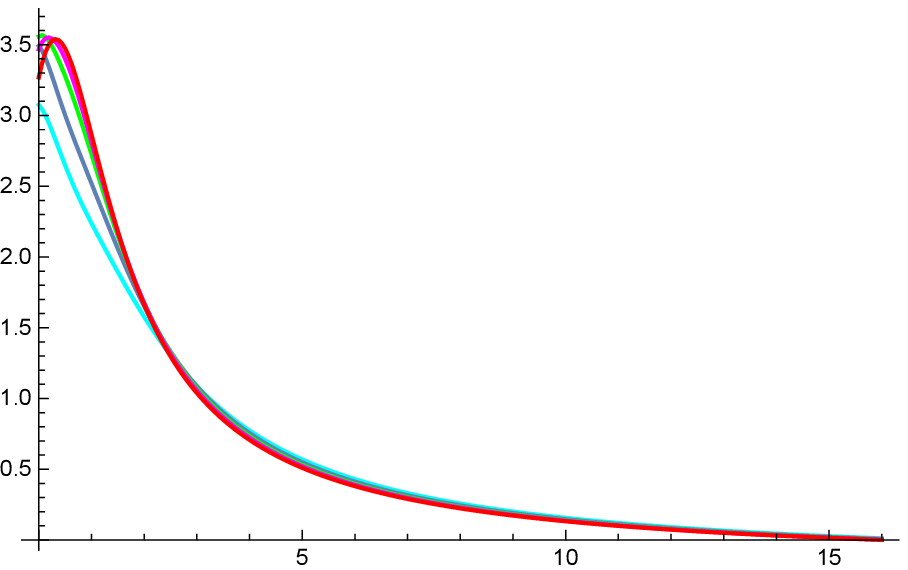} 
\end{center}
\end{minipage}
\hfill
\begin{minipage}[b]{0.47\linewidth}
\begin{center}
\includegraphics[width=7cm,angle=0,clip]{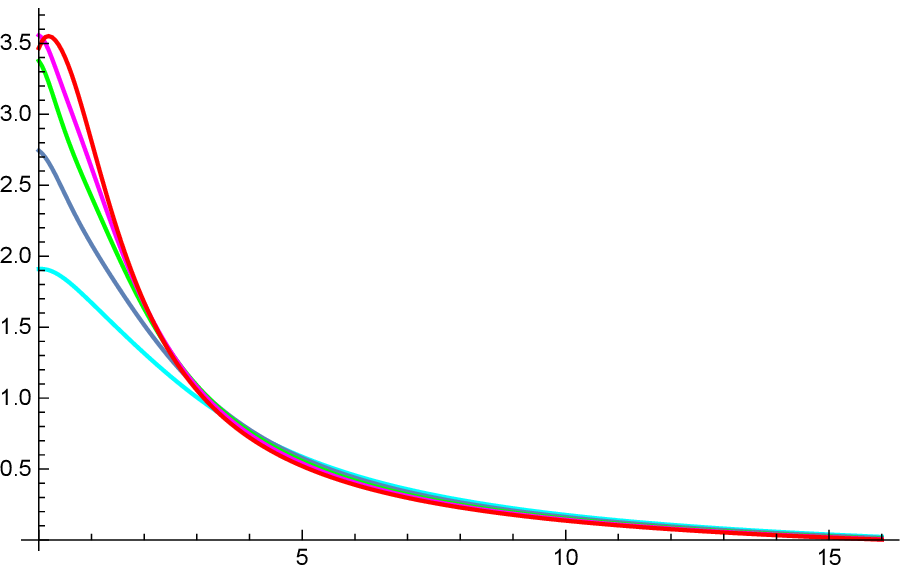} 
\end{center}
\end{minipage}
\caption{ A type MC plaquette action of $u_2=\frac{N_{even}}{16}$(left) and $u_2=\frac{N_{odd}}{16}\Delta u$(right). The sample $\alpha$. }
\end{figure*}

\begin{figure*}[htb]
\begin{minipage}[b]{0.47\linewidth}
\begin{center}
\includegraphics[width=7cm,angle=0,clip]{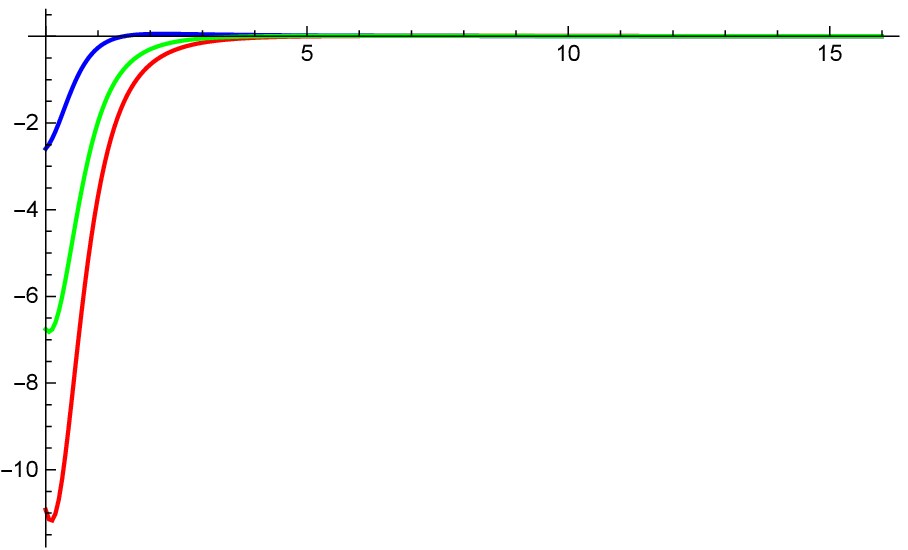}  
\end{center}
\end{minipage}
\hfill
\begin{minipage}[b]{0.47\linewidth}
\begin{center}
\includegraphics[width=7cm,angle=0,clip]{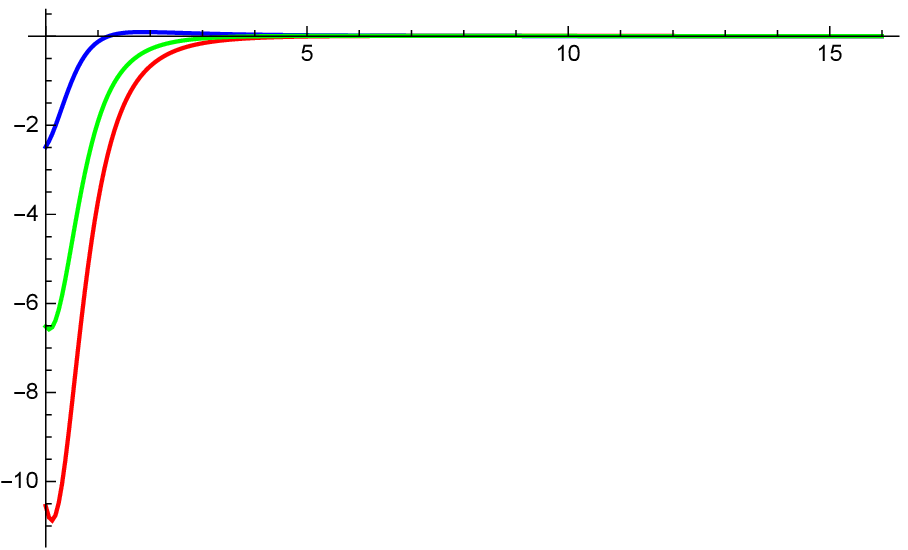}
\end{center}
\end{minipage}
\caption{ B type MC link actions of $u_2=0$(left) and $u_2=\frac{1}{16}\Delta u$ (right). $f$link (red), $e$ link(blue) and the average(green) of the sample $\gamma$.}
\end{figure*}
\begin{figure}[htb]
\begin{center}
\includegraphics[width=7cm,angle=0,clip]{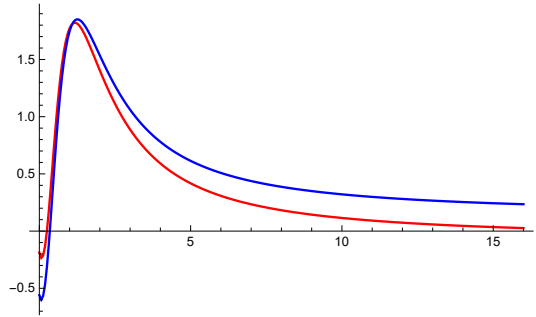}  
\end{center}
\caption{B type MC [Plaquette+ ($e$-link(blue)+ $f$- link(red))/2] actions of $u_2=0$(red) and $u_2=\frac{1}{16}\Delta u$(blue).}
\end{figure}
\begin{figure}[htb]
\begin{center}
\includegraphics[width=7cm,angle=0,clip]{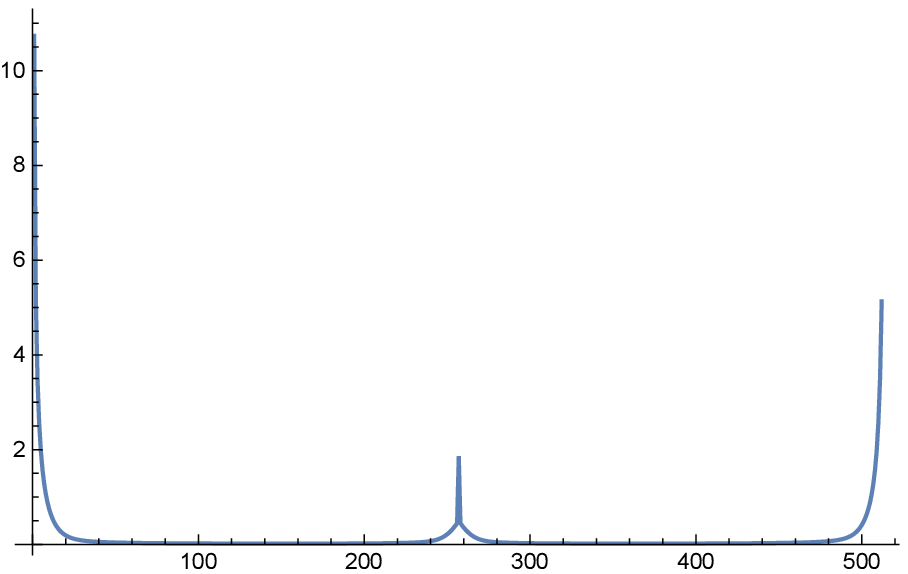}
\end{center}
\caption{Fourier transform of $u_2=\frac{(N_{even}+N_{odd})}{16}\Delta u$ total actions.}
\end{figure}
\section{Perspective of detecting hysteresis effects using groupoids }
For understanding the NDT method, it is useful to study the hysteresis produced by the nonlinearity of the fermion interactions and noncommutativity of quaternions\cite{Chevalley46,Lounesto01,Madore95} which represent the Weyl fermions. 

Taking into account the property of convolution $\delta_a*\delta_b=\delta_{a+b}$, convolutions of an ordinary wave and the TR wave are expected to yield a peak $\delta_0$, if the wave propagations are uniform. On a 2D projected space, the B-type loops, the actions $S^e_{LB}$ and $S^f_{LB}$ were in general quite different. 

 In order to analyze the path dependence, we adopt the notion of groupoid appeared in the framework of
cathegory and functor in the book of Mac Lane \cite{MacLane86}. A groupoid is a generalization of a group. The homotopy classes of paths in a topological space form a category under composition of paths. A category in which every arrows are invertible is called a groupoid. 

Application of groupoid in quantum mechanical von Neumann algebra \cite{vonNeumann32} was done by Connes\cite{Connes94}, and reviewed by Weinstein\cite{Weinstein96} as a tool for unifying internal and external symmetry.

Algebraic structures of groupoid $G$ and its distinguished subset $G^{(0)}$ is written in the form
\[
(a*b)(\gamma)\sum_{\gamma_1\circ \gamma_2=\gamma}a(\gamma_1)b(\gamma_2)
\]
where $\gamma\in G$ and
\[
\circ:G^{(2)}=\{(\gamma_1,\gamma_2)\in G\times G; s(\gamma_1)=r(\gamma_2)\}\to G
\]
where the source map $s$ and the range map $r$ in $G$ satisfy
\begin{eqnarray}
(1)&& s(\gamma_1\circ\gamma_2)=s(\gamma_2), r(\gamma_1\circ\gamma_2)=r(\gamma_1).\nonumber\\
(2)&&s(x)=r(x)=x, \quad x\in G\to G^{(0)}\nonumber\\
(3)&&\gamma\circ s(\gamma)=\gamma, r(\gamma)\circ \gamma=\gamma\nonumber\\
(4)&&(\gamma_1\circ\gamma_2)\circ\gamma_3=\gamma_1\circ(\gamma_2\circ\gamma_3)\nonumber\\
(5)&& \gamma\gamma^{-1}=r(\gamma), \gamma^{-1}\gamma=s(\gamma)\nonumber
\end{eqnarray}

We consider $G^{(0)}\subset M\times[0,1]$ with inclusion
\begin{eqnarray}
&&(x,\epsilon)\to(x,x,\epsilon)\in M\times M\times ]0,1]\mbox{  for } x\in M,\epsilon>0 \nonumber\\
&&(x,0)\to x\in M\subset TM\nonumber
\end{eqnarray}
and the range map and the source map
\[
\left\{ \begin{array}{rl}
r(x,y,\epsilon)=(x,\epsilon) \mbox{ for } x\in M,\epsilon>0 \\
         r(x,X)=(x,0)\mbox{ for } x\in M, X\in T_x(M). 
         \end{array}
         \right. 
 \]
 \[        
 \left\{ \begin{array}{rl}
s(x,y,\epsilon)=(y,\epsilon) \mbox{ for } y\in M,\epsilon>0 \\
      s(x,X)=(x,0)  \mbox{ for } x\in M, X\in T_x(M). 
         \end{array}
         \right.\nonumber
 \]
          
The composition is given by
\begin{eqnarray}
&&(x,y,\epsilon)\circ(y,z,\epsilon)=(x,z,\epsilon) \mbox{ for } \epsilon>0 \mbox{and }x,y,z\in M.\nonumber\\
&&(x,X)\circ(x,Y)=(x,X+Y) \mbox{for } x\in M \mbox{ and } X, Y\in T_x(M).\nonumber
\end{eqnarray}
where $T_x(M)$ is the tangential plane of the manifold $M$.

We apply the composition to a system with hysteresis and without hysteresis governed by instantons\cite{BBHN96}. 

Instanton hypothesis is to take the tangential plane at $A$ and at $B$ represented by dashed lines in Fig.\ref{holonomy} are prallel $T_x(M)$. $x$ is taken at $A$ and $y$ is taken at $B$.

When we consider the hysteresys, $x$ is taken at $A'$ and $y$ is taken at $B$, but the tangential plane at $B$ is chosen to be orthogonal to the Pluecker vector $CC'$\cite{Yokonuma97}.
The tangential vector $X$ is expressed as
\[
x_n\to x, y_n\to y, \frac{x_n-y_n}{\epsilon_n}\to X.
\]

 Instead of instanton configurations of the spin system, we adopt the fixed point (FP) actions of block spins defined by quaternions
\[
{\bf S}=S^1(u_1,u_2) e_1+S^2(u_1,u_2) e_2+S^0(u_1,u_2) e_1 e_2,
\]
where $(u_1,u_2)$ are coordinates inside Wilson loops, that defines the minimal action of the system.

 The FP action is a subset of the 4D Wilson action of De Grande et al. \cite{DGHHN95}. The Dirac spinors are replaced by Weyl spinors.

 The Trotter product formula says that for finite dimensional matrices $A,B$
\[
exp(A+B)=\lim_{n\to \infty}[exp(A/n) exp(B/n)]^n
\]
When $A,B$ are self adjoint and $A+B$ is self adjoint in $D(A)\cap D(B)$
\[
s-lim_{n\to\infty}(e^{\sqrt{-1}t A/n} e^{\sqrt{-1}t B/n})^n=e^{\sqrt{-1}(A+B)t}
\]
The semi-group $T_t$ is defined as an operator that satisfy
\[
T_t T_s=T_{t+s}
\]
and for an element in Banach space $X$, continuity condition
\[
\lim_{t\to 0}T_t f=f,\quad f\in X
\]
is satisfied.

There are at least three definitions of grupoids\cite{SW96}. In a definition of non-gauge symmetic tiling\cite{Petitjean21b}, the author criticized \cite{Weinstein96} and  claimed that local symmetries can be defined as global symmetries.  His conclusion comes from his choice of the group $G=\{e,z,z^2=\bar z\}$ and exclusion of $z=\bar z$ case.\cite{Petitjean21a}. Connes considered in non-commutative geometry, the group $\{e, z,z^*\}$ with self-adjoint projection $e=e^*=e^2 $ and its equivalent projection $f\sim e$ and equivalent action $\alpha_t'(f)=f$ for $t\in {\bf R}$, 

 In our case, the global symmetry is not defined by the local symmetry since the local tangential vector at $A$ and $B$ are not unique and so the global symmetry is not defined by a local symmetry. The global Gauss-Bonnet theorem cannot trivially derived from the local Gauss-Bonnet theorem\cite{doCarmo76}.

 Saito defined $H_0=-c_0^2\Delta$ and $B_0=\sqrt{H_0}$, $\varphi(t)=\langle u(t),u_t(t)\rangle$,
\begin{eqnarray}
&&||\langle u,v\rangle||^2=||B_0 u||^2_2+||v||^2,\nonumber\\
&&A_0=\sqrt{-1}\left(\begin{array}{cc}
                                0&I\\
                                -B_0^2&0\end{array}\right), \quad
                                D(A_0)=D(B_0^2)\oplus D(B_0) \nonumber\\
&&\varphi'(t)=-\sqrt{-1} A_0\varphi(t), \quad \varphi(0)=\varphi_0\nonumber\\
&&\varphi(t)=W_0(t)\varphi_0.\nonumber
\end{eqnarray}
\[
W_0(t)=\left(\begin{array}{cc}
               \cos B_0 t& B_0^{-1}\sin B_0 t\\
               -B_0\sin B_0 t&\cos B_0 t\end{array}\right)
\]

 And $H_1=-c(x)^2\rho(x)\nabla\cdot \rho(x)^{-1}\nabla$, and $B_1=\sqrt{H_1}$.
\begin{eqnarray}
&&A_1=\sqrt{-1}\left(\begin{array}{cc}
                                0&I\\
                                -B_1^2&0\end{array}\right), \quad
                                D(B_1)=D(B_1^2)\oplus D(B_1) \nonumber\\
&&\varphi(t)=W_1(t)\varphi_0,\nonumber\\
&&W_1(t)=\left(\begin{array}{cc}
\cos B_1 t& B_1^{-1}\sin B_1 t\\
-B_1\sin B_1 t&\cos B_1 t \end{array}\right)\nonumber
\end{eqnarray}

 We consider $t$ dependence through the propagation of a wave packet $\varphi_0$ via $t-x/c_0$ term.
The spatial function $\varphi_0$ in derived from the Fourier transform of momentum space wave function $\hat\varphi(u_1,u_2)$ on 2 dimensional(2D) lattices with torus structure $T^2$. 

 The Topological Dynamics on $T^2$ is explained in the textbook of Saito \cite{Saito71}.
 Dynamics of orbits on  $S^2$ was studied by Poincar\'e \cite{Poincare99} and generalized to orbits on torus $T^n (n>2)$ by Weyl. On $S^2$, the minimum set is a point or a curve and not dense in $S^2$, but on $T^2$, it becomes dense.

We consider $x=(x_1,x_2)$ as a path in a manifold $X$ and $f=(f_1, f_2)$ as a funtion defined in $X=T^2$:
\[
\frac{dx_1}{dt}=f_1(x_1,x_2), \quad \frac{dx_2}{dt}=f_2(x_1,x_2),
\]
that satisfy $\frac{\partial f_1}{\partial x_1}+\frac{\partial f_2}{\partial x_2}=0$.

 $f_1$ and $f_2$ can be expanded by Fourier series
\begin{eqnarray}
f_1&=&a^1+\sum_{m,n} a^1_{mn}\cos 2\pi(m x_1+n x_2)\nonumber\\
&&+\sum_{m,n}b^1_{mn}\sin 2\pi(m x_1+n x_2),\nonumber\\
f_2&=&a^2+\sum_{m,n} a^2_{mn}\cos 2\pi(m x_1+n x_2)\nonumber\\
&&+\sum_{m,n}b^2_{mn}\sin 2\pi(m x_1+n x_2),\nonumber
\end{eqnarray}

 When $a^1:a^2$ is rational, orbits are periodic. When it is irrational, orbits are dense in $T^2$. In addition to the plus orbitally stable solutions $L^+(x)$, the minus orbitally stable solutions $L^-(x)$, there are 
wandering solutions $W(x)$, which are defined for a neighbourhood $U$ around $x$, $\pi(U,t)\cap U=\phi$ for $t>\tau$, where $\tau$ is a sufficiently large number.

$t\to \infty$: $L^+(x)$ and at $t\to -\infty$: $L^-(x)$.
When $C(x)\cap L^{+/-}(x)=\phi$, $C(x)$is called positively/negatively Poisson stable,  and when both relations are satisfied $C(x)$ is called Poisson stable.

 A solution with $x=\xi$ at $t=0$ is written as $x=\pi(\xi,t)$. 
Saito defined
 \[
 C^+ (\xi,t)=\cup_{0\leq t<\infty}\pi(\xi,t),\quad
 C^- (\xi,t)=\cup_{-\infty<t\leq 0}\pi(\xi,t)
 \]
 and  $C(\xi)=\cup_{-\infty<t<\infty}\pi(\xi,t)=C^+ (\xi)\cup C^- (\xi)$.

 Dynamical systems are given by triplet $(X,{\bf R},\pi)$. We consider $X=T^2$ a 2D torus.
 Schroedinger methods of classical acoustics contain spinor structure.
 
 Saito defined a sphere of radius $\epsilon$ with center $x$: $S(x,\epsilon)=[y\in M; d(x,y)<\epsilon]$, 
where $M$ is a set of closed Poisson stable orbits $\overline{C(x)}$, and define $F_k$ as a set of $x$
such that $\pi(x,t)\cup S(x,\epsilon_k)=\phi$ for $t>t_k$. Similarly $F_k^*$ is defined as a set of $x$ such that
$\pi(x,t)\cup S(x,\epsilon_k)=\phi$ for $t<-t_k$. He prooved that $M-(\cup F_k)\cup(\cup F_k^*)$ is perfectly separable.

When for an arbitrary $\epsilon>0$, a positive number $L(x)$ is defined such that for  $|\tau|\geq L(\epsilon)$
\[
C(x)\subset S(x(t,t+\tau),\epsilon)
\]
is satisfied for all $t$, the point $x$ is called recurrent. 

It was shown that a Poisson stable minimal set is a closure of recurrent orbits.  

 Birkhoff\cite{Birkhoff27} separated $X=X_1+W_1$, where $X_1$ is a closed invariant set, while $W_1$ is an open invariant set.

 In the dynamical system $(X_1,{\bf R},\pi)$, $X_1$ can be separated as $X_2+W_2$. We can extend the analysis to $(X_\alpha,{\bf R},\pi)$ where $X_\alpha$ is a central set, which contains complicated orbits.

 The Maupertuis' minimal action principle says 
\[
J=\int_{t_0}^{t_1}(-F(x_1,x_2)+\sum _i y_i\frac{dx_i}{dt})dt
\]
has zero variation when $\frac{dx_i}{dt}=\frac{\partial F}{\partial y_i},$ $\frac{dy_i}{dt}=-\frac{\partial F}{\partial x_i},$
provided $\delta x_i$ at $t_0$ and $t_1$ are zero.

Hamilton's minimal action is given by changing $F$ to $F-h$ where $h$ is the energy integral.

 The phase propagation and the translation in the position space induced by Hamiltonians leads to search of orbits defined by the Lie group.
We consider $O(3)$ nonlinear $\sigma$ model in 2 dimentional Euclidean space proposed by Blatter et al.\cite{BBHN96} for all values of $s$ and $t$, $U_{a,t}=T_t T'_{at}$ where $a$ is a positive constant, is also a semi-group. 

When they do not commute, 
\[
S_{a,t}=\lim_{h\to 0}(T_h T'_{ah})^{[t/h]}
\]
where $[t/h]$ is the greatest integer in $t/h$, become a semi-group.

 Let $X$ be continuous function space on the real line, and $T_t$ and $T'_t$ be semi-group of left and right translations:
$T_t f(x)=f(x-t)$, $T'_t f(x)=f(x+t)$, their domain $D(T)=\Omega$, $D(T')=-\Omega$

 Connes defined Groupoids as a tool for producing the semidirect product of a local Lie group $G$, and a map $\Gamma$ of source to range or vice-versa. 

 The noncommutative geometry is an extension of the Heisenberg equation of motion $dx=[H,x]$ to
$df=[F,f],\in {\mathcal L}^{n+1}({\mathcal H})$, where $n$ is an integer.
 Chernoff's theorem for self-adjoint operator $0\leq f(t)\leq 1$ says that when $S(t)=t^{-1}(1-f(t))$ converges in strong resolvent set to $A$ as $t\to 0$
\[
s-\lim_{n\to\infty}f (\frac{t}{n})^n=e^{-tA}.
\]
 The graph of a linear transformation $T$ is the set of pairs $\{\langle \varphi,T\varphi\rangle|\varphi\in D(T)\}$. The graph $\Gamma(T)$ is a subset of the Hilbert space ${\mathcal H}\times {\mathcal H}$.

If $\Gamma(T_1)\supset \Gamma(T)$, $T_1$ is an extension of $T$.  An operetor $T$ is closable, if it has a closable extension.

 A family $\{T(t)|0\leq t\leq \infty\}$ of bounded operators on a Banach space $X$ is called a contraction semi-group if a) $T(0)=I$, b) $T(s)T(t)=T(s+t)$ for all $s,t\in {\bf R}^+$, c) For each $\varphi\in X$, $T\to T(t)\varphi$ is continuous.

 Let $A$ and $B$ be generators  of contraction semi-groups on a Banach space $X$, the closure of $A+B$ denoted as $\overline{A+B}$ satisfies 
  \[
 e^{-t(\overline{A+B})}\varphi=\lim_{n\to\infty}(e^{-tA/n} e^{-tB/n})^n \varphi
 \]
 on $D=D(A)\cup D(B)$.
 
 We considered one loop fixed point actions, and quaternion Borel Transform, and discussed Gauss-Bonnet topological characteristics that could be measured in time-reversal conserving, rotational symmetry violating phonon propagations. 
 
For a start vector $A_s$ and its end vector $A_e$ via transformation on a link, gravitational shift shown in Misner et al \cite{MTW70} is given by Riemann normal coodinates as 
\[
\delta A=e_\lambda\wedge e_\mu R^{|\lambda\mu|}_{xy}\Delta x\Delta y,
\]
where $\lambda,\mu$ in $R^{|\lambda\mu|}_{12}$ are summed with the restriction $\lambda<\mu$. 

Since $R^{03}_{12}=-R^{03}_{21}$, link actions of A-type loops are expected to cancel.

In B-type loops, since we can take $e_1\wedge e_2\sim e_t$, the hysteresis and holonomy group play essential roles. 

We considered one loop fixed point actions, and quaternion Borel Transform, and discussed Gauss-Bonnet topological characteristics that could be measured in time-reversal conserving, rotational symmetry violating phonon propagations. 

In $(2+1)D$ system, renormaizability of Gaussian Borel transform is proved by Koch and Wittwer\cite{KW86,KW91,KW94}, and properties of the infrared fixed point were investigated. 
They showed that when $d\leq 3$ the renormalization group series converges.
 
\section{Summary and Conclusion}
 Gravitational effects could be incorpolated as a thermal background effects.  Luescher's domain decomposition method\cite{Luescher03,Luescher86} for lattices with boundary matches Clifford algebra using real quaternions, since one can take Clifford pairs on the boundary with Gaussian average zero, and asymptotically free distributions. 

The MC samples $\alpha,\beta,\delta$ yield Poisson stable $L^+(x)$ and $L^-(x)$. The sample $\gamma$ yields the recurrent Lyapunov stable minimal set.   

 MC simulations on a $256\times 256$ lattice revealed $u_1$ spectrum depends on $u_2=\frac{N}{16}\Delta u$ whether $N$ is even or odd. Singular behaviors observed in $128\times 128$ are due to mixing of two boundary conditions of coarse lattices. In NDT, clean signals would be obtained when $e$ link action and $f$ link action have similar magnitude of shifts, since plaquette action becomes asymptotic free when $N$ is even.

 In comparison of NDT signal and the MC data, machine learning or neural network technique may be applicable\cite{Aggarwal18}.

The Clifford group projects the wave functions invariant under Levi-Civita parallel transformations.

If our assumption is right, properties of phonetic solitons will reveal gravitational effects\cite{RML12,Adler86} on the infrared fixed point.

\begin{acknowledgments}
SF thank the RCNP and CMC of Osaka University for allowing use of super computers there, and giving informations on parallel computations, and Prof. M. Arai at Teikyo Univ. for supports.
\end{acknowledgments}


\end{document}